\documentclass[preprint,12pt]{aastex}

\begin{document}

\shorttitle{{\it Swift}'s Burst Sensitivity}

\title{Post-Launch Analysis of {\it Swift}'s Gamma-Ray Burst Detection Sensitivity}
\author{David~L.~Band\altaffilmark{1,2}}
\altaffiltext{1}{Code 661, NASA/Goddard Space Flight
Center, Greenbelt, MD  20771}
\altaffiltext{2}{Joint Center for Astrophysics, Physics Department,
University of Maryland, Baltimore County, 1000 Hilltop Circle,
Baltimore, MD 21250}
\email{dband@milkyway.gsfc.nasa.gov}

\begin{abstract}
The dependence of {\it Swift}'s detection sensitivity on a burst's
temporal and spectral properties shapes the detected burst
population. Using simplified models of the detector hardware and the
burst trigger system I find that {\it Swift} is more sensitive to
long, soft bursts than {\it CGRO}'s BATSE, a reference detector
because of the large burst database it accumulated. Thus {\it Swift}
has increased sensitivity in the parameter space region into which
time dilation and spectral redshifting shift high redshift bursts.
\end{abstract}


\section{Introduction}

The gamma-ray bursts which {\it Swift} detects depend on the
physical properties of the Burst Alert Telescope (BAT), {\it
Swift}'s gamma-ray detector, and on the BAT's triggering system. The
dependence of the BAT's detection sensitivity on a burst's temporal
and spectral characteristics shapes the burst population that {\it
Swift} studies.  While the {\it Swift} observations are revealing a
wealth of new phenomena through the study of individual bursts, we
also want to relate these bursts to the bursts studied by previous
missions.  In particular, because of the large and statistically
well-defined sample of more than 2700 bursts it collected, the Burst
and Transient Source Experiment (BATSE) on the {\it Compton
Gamma-Ray Observatory (CGRO)} is the reference detector to which
subsequent detectors such as the BAT are compared. Therefore, in
this work I use the BAT's on-orbit calibration to gain a deeper
understanding of the detector's sensitivity to different types of
bursts. The insight from this study will help the design of future
missions, such as {\it EXIST} (Grindlay 2005).

The BAT detects approximately 100 bursts per year. Compared
to BATSE's burst sample, a higher fraction of the bursts
the BAT detects are long duration ($T_{90}>2$~s) bursts
(see Figure~1), although the few short duration bursts that
have been detected have been particularly revelatory.
Understanding the observed duration distribution is a goal
of this work.

Understanding the BAT's burst detection sensitivity requires the
sequence of events on board the spacecraft. Bursts are detected by
the BAT (Barthelmy et al. 2005), a large field-of-view
(FOV---1.4~sr), 15--150~keV coded mask detector with a 5200~cm$^{2}$
cadmium-zinc-teluride (CZT) detector plane.  The BAT's detector
plane is sensitive to higher energy photons, but burst imaging and
spectroscopy have an effective high energy cutoff of $\sim 150$~keV.
Once the BAT detects a burst, the spacecraft slews autonomously
(within operational constraints) to place the burst location in the
center of the much smaller FOVs of the X-ray Ray Telescope
(XRT---Osborne et al. 2005) and the coaligned UV-Optical Telescope
(UVOT---Roming et al. 2005). Thus the BAT's trigger system
determines which bursts are detected, although the other two
detectors' performance and operational constraints affect whether
the afterglow is followed by {\it Swift} immediately after the
burst.

The BAT's flight software detects bursts on board in two steps
(Fenimore et al. 2003, 2004; Palmer et al. 2004). A rate trigger
monitors the count rate from the CZT detectors for a statistically
significant increase; the BAT's rate trigger is complex, testing the
count rate from the detector plane (and subsections of the plane) on
timescales ranging from 0.004~s to 32~s using a variety of different
background estimates. Once a rate trigger occurs, an image is formed
through the coded mask system. To maximize the signal-to-noise ratio
of the counts used for imaging, the software varies: the energy
band; the `foreground' time period over which the counts are
accumulated; and the `background' time periods before, and perhaps
after, the foreground time period from which the background during
the foreground period is estimated. A number of images may be formed
using different foreground time periods before the significance
exceeds a detection threshold or the software concludes that a
detection is not possible. Only if a new statistically significant
point source is evident in an image is a burst considered to be
detected. Periodically (once every 64 and 320~s, and when the
spacecraft changes its orientation---Palmer et al. 2004; McLean et
al. 2004) an image is formed and checked for a new point source even
without a rate trigger. Because a burst detection requires the
imaging of a new point source, the threshold for the rate trigger is
set to permit many false positives that are subsequently rejected by
the imaging step. Consequently, the imaging step is usually the most
restrictive step and therefore determines the BAT's burst
sensitivity.

The BAT's trigger system is complex with many triggers and
background estimates (Fenimore et al. 2003, 2004; Palmer et al.
2004).  The flight software turns triggers on and off based on the
computational load.  While diagnostics are telemetered to the
ground, the telemetry stream cannot provide sufficient data to
reproduce on the ground the behavior of the trigger system precisely
at all times. The complexity of the trigger system maximizes the
BAT's sensitivity---{\it Swift}'s design goal---at the expense of
making an accurate determination of this sensitivity at a given time
very difficult if not impossible.  In particular, the BAT achieves
high image sensitivity by accumulating counts over much longer
timescales than BATSE did, making the trigger sensitive to the
details of the burst lightcurve. In contrast to burst spectra whose
shapes are adequately described by two or three parameters,
lightcurves differ greatly from burst to burst when considered on
timescales greater than a second, and cannot be parameterized for
sensitivity calculations by only a few parameters.

Despite all these caveats, I develop a semi-quantitative
understanding of the burst populations that the BAT detects using a
simplified model of the BAT's trigger system that captures the
essential features of the trigger.  I assume that imaging is the
most restrictive step of the trigger, and therefore ignore the
complexity of the rate trigger.  I use a signal-to-noise ratio
estimate of the sensitivity of the imaging step.  This calculation
captures the fundamental dependence of the BAT's sensitivity on a
burst's hardness and duration.  Calculations with greater
verisimilitude would result from applying the BAT trigger code to
simulated burst data; the BAT team maintains a working copy of the
trigger code incorporated in the flight software. Before launch such
simulations were run to verify the performance of the BAT and its
flight software (Fenimore et al. 2004) and to determine the trigger
system's initial settings (McLean et al. 2004). The simulated bursts
should be accurate representations of the bursts the BAT might
detect.

I separate my evaluation of burst sensitivity into the dependencies
on the burst's spectrum and lightcurve.  This is an approximation,
since a burst's spectrum changes during the burst---usually the
spectrum softens with time (Ford et al. 1995)---and the lightcurve
depends on the energy band---usually individual pulses and the
duration of the entire burst are shorter at high energy. After first
providing the formulae for the detection significance for rate and
image triggers (\S 2.1), I evaluate the BAT's energy-dependent (\S
2.2) and duration-dependent (\S 2.3) burst sensitivity. I use these
results to understand the observed burst population (\S 3). While I
have discussed the factors affecting the BAT's sensitivity with the
members of the BAT instrument team, the conclusions are my own.  I
use preliminary values for the performance of the instrument and the
mission, and thus my calculated sensitivities utilizing a simple
model of the BAT trigger should be regarded as illustrative, not
definitive.

\section{The BAT's Burst Sensitivity}

\subsection{Burst Triggers}

While they involve very different operations, rate and imaging
triggers both analyze the counts accumulated by a burst detector
over an energy band $\Delta E$ and accumulation time $\Delta t$; the
BAT's flight software analyzes overlapping energy bands (see \S 2.2)
and accumulation times (see \S 2.3). For the BAT, the initial rate
trigger and the image with a statistically significant point source
detection need not use the same $\Delta E$ and $\Delta t$. If a
burst is present then the number of observed counts in $\Delta E$
and $\Delta t$ is the sum of the counts from the burst $C_s$ and the
background $B$.

Rate and imaging triggers have similar dependencies on source and
background counts, and thus analogous methods can be used to
evaluate the resulting sensitivities. Before {\it HETE-II} and the
BAT most burst detectors, such as BATSE, used
\begin{equation}
S_r = {{C_s}\over \sqrt{B}}
\end{equation}
as the detection significance for a rate trigger: the increase in
the number of counts over the background is compared to the
background's fluctuation scale.  For a trigger, $S_r$ must exceed a
threshold value.  To mitigate difficulties that occur when $B$ is
very large or very small, the BAT's rate trigger replaces the
background $B$ in the denominator of eq.~1 with a sum $D$ of terms
(see eq.~3 of Fenimore et al. 2003). When $B$ is small $D$
asymptotes to a constant, while $D$ asymptotes to $B^2$ when $B$ is
large, converting the detection criterion from a signal-to-noise
ratio to a signal-to-background ratio. For intermediate values of
$B$, $D$ is approximately equal to $B$.

The significance of a point source in a coded mask image is
(G.~Skinner 2005, personal communication)
\begin{equation}
S_i = {{f_m C_s}\over \sqrt{C_s+B}}
\end{equation}
where the factor $f_m$ compensates for the finite size of
the detector pixels relative to the mask elements.  One
interpretation of this factor of $f_m$ is that the finite
size of the detector pixels smears images on the sky,
thereby lowering their significance.  Skinner (2005,
personal communication) finds $f_m=0.73$ for the BAT, which
explains why the rate trigger significance is greater than
the image significance for the bursts the BAT detects
(D.~Palmer 2005, personal communication).  The fluctuation
level for an image includes the source counts in addition
to the background counts (i.e., the denominator in eq.~2 is
$\sqrt{C_s+B}$ and not $\sqrt{B}$) because the burst counts
are merely background for positions on the sky other than
that of the burst (the origin of $C_s$).

The rate trigger significance $S_r$ in eq.~1 is applicable
to BATSE but not to the BAT's rate trigger for which the
significance differs significantly from $S_r$ in eq.~1 for
both small and large numbers of background counts. However,
imaging almost always is the most restrictive step in the
BAT's burst detection algorithm, and therefore I only
consider the sensitivity resulting from imaging.  In
comparing the BATSE and BAT sensitivities, I use eq.~1 for
BATSE and eq.~2 for the BAT.  I also assume that the BAT
flight software successfully finds the `foreground' time
period (the time period used for imaging) that maximizes
the signal-to-noise ratio, thereby optimizing the image
step.

In the analysis that follows I first consider the energy dependence
of $S_i$ holding the accumulation time $\Delta t$ fixed at 1.024~s
(\S 2.2), a background-dominated case that allows me to use the
methodology developed for rate triggers (Band 2003). Subsequently I
consider the dependence on burst duration (\S 2.3).

\subsection{Energy Dependence}

In this subsection I assume that $\Delta t$=1.024~s, for
which the background dominates the burst counts at
threshold (i.e., $B\gg C_s$).  I calculate the number of
source counts $C_s$ by convolving the burst spectrum with
the effective area over the energy band $\Delta E$.
Figure~2 shows the current understanding of the BAT's
detector efficiency (D.~Hullinger, 2005, personal
communication), here defined as the effective area on-axis
divided by {\it half} of the area of the detector plane
(the detector plane area is divided by two to account for
the coded mask); thus this efficiency is equivalent to the
efficiency for a detector with half the area and no mask.

Ideally the closed cells of the mask (consisting of lead
tiles) would be perfectly opaque, but at high energy (above
$\sim$100~keV) the optical depth through the lead tiles
decreases.  Imaging with a coded mask relies on the shadow
cast by the closed mask cells. However, if flux leaks
through the closed mask cells, then the contrast between
the detector pixels that are illuminated by the source and
those that are shadowed is reduced; this leakage is
equivalent to no flux leaking through the closed mask cells
and the detection of less flux by the illuminated detector
pixels.  The solid curve on Figure~2 is the net detector
efficiency, the efficiency for the difference between the
fluxes through the open and closed mask elements, which is
relevant for imaging.  Because the imaging step is the most
restrictive part of the BAT's trigger, the net detector
efficiency is used to calculate the source counts for the
sensitivity.

The dashed curve on Figure~2 is the gross detector efficiency, the
efficiency for the sum of the fluxes through the open and closed
mask elements.  The product of the gross detector efficiency, the
incident flux, the total detector area, and the fraction of the
coded mask that is open (half for the BAT) results in the total
count rate.  The gross detector efficiency is relevant for the BAT's
rate trigger.

Thus the net detector efficiency is reduced at the energies where
the mask's lead tiles are partially transparent while the gross
detector efficiency is increased at the same energies because more
source photons reach the detector plane.  Note that the rate trigger
and imaging use different detector efficiencies and will have
somewhat different energy dependencies.

CZT has high quantum efficiency below 100~keV.  However the optical
depth through the mask substrate that supports both the closed and
open mask cells results in the roll-off in the detector efficiency
at low energy ($<$40~keV). This low energy roll-off reduces the
aperture flux (i.e., the cosmic X-ray background), which dominates
the total background at low energy, but also decreases the BAT's
sensitivity to X-ray Flashes and X-ray rich bursts.

The BAT's total on-orbit background rate is approximately
$\sim$10~kHz, consistent with pre-launch estimates.  At low energies
the background is dominated by the aperture flux---the cosmic X-ray
background through the mask---while at high energy instrumental
background and aperture flux through the BAT's side shields (which
become transparent at $\sim 100$~keV) increase the background.  The
background varies over an orbit, and I use the lowest observed
background rates in the energy bands used by the BAT's burst
trigger: $\sim$2300~Hz for $\Delta E=$15--25~keV; $\sim$4700~Hz for
$\Delta E=$15--50~keV; $\sim$4700~Hz for $\Delta E=$25--100~keV; and
$\sim$4700~Hz for $\Delta E=$50--500~keV.  The background is near
this minimum about half the time; the reduction in sensitivity
resulting from higher backgrounds reduces the overall burst
detection rate by about 5\%.  Note that for $B=4700$ counts in 1~s,
$C_s\sim S_{i,{\rm th}}\sqrt{B}/f_m\sim 700$, and thus $C_s \ll B$
at a threshold value of $S_{i,{\rm th}}=7$; the assumption that the
background dominates is valid.

I parameterize the spectrum with the `Band' function (Band
et al. 1993) which is a smoothly broken power law with low
energy spectral index $\alpha$ ($N(E)\propto E^\alpha$) and
high energy spectral index $\beta$ ($N(E)\propto E^\beta$).
The characteristic energy is the peak energy $E_p$, the
photon energy of the peak of the $E^2N(E)\propto \nu f_\nu$
spectrum.  The spectrum can be normalized by the flux
integrated over a specified energy band, which need not be
the same as $\Delta E$.  I use $F_T$, the peak flux in the
1--1000~keV band.

Figure~3 compares the maximum sensitivity for BATSE (left) and the
BAT on-axis (right) for $\Delta t$=1.024~s.  The sensitivity is the
threshold peak photon flux $F_T$ at which the detector triggers.
Because the threshold is expressed in the same units---the flux in
the 1--1000~keV band---regardless of $\Delta E$, the sensitivity of
different detectors and of different $\Delta E$ for the same
detector, can be compared (Band 2003).  These thresholds depend on
the burst's spectral parameters that determine the shape of the
spectrum (not its normalization). The curves on Figure~3 show the
threshold flux as a function of $E_p$, holding the low and high
energy spectral indices $\alpha$ and $\beta$ fixed. The BAT runs its
trigger on four different $\Delta E$ simultaneously ($\Delta
E=$15--25~keV, 15--50~keV, 25--100~keV, and 50--500~keV), and the
detector sensitivity is the lowest threshold at any given $E_p$,
resulting in the scalloping of the BAT sensitivity curves.

As can be seen, on-axis the BAT is less sensitive than BATSE's
maximum sensitivity for $E_p>100$~keV by a factor of $\sim$1.5, and
is more sensitive at lower $E_p$ values, again by a factor of
$\sim$1.5, for the same $\Delta t=1.024$~s.  As will be discussed
below, the BAT's overall sensitivity depends on both its sensitivity
at fixed $\Delta t$ and the sensitivity resulting from triggering on
multiple values of $\Delta t$.

\subsection{Duration Dependence}

I now consider the BAT's sensitivity to bursts with
different durations.  BATSE used a rate trigger with three
values of $\Delta t$---0.064, 0.256 and 1.024~s---while
after a rate trigger the BAT can form images on a variety
of timescales ranging from $0.004$~s to $26$~s.  In
addition, the BAT forms images every 64 and 320~s without a
rate trigger.

The relationship between burst duration and the detector
accumulation time $\Delta t$ is illustrated by considering a
constant flux burst of duration $T$ when the background dominates
the source. If $T > \Delta t$, that is, if the flux remains constant
over $\Delta t$, then the threshold flux is proportional to $\Delta
t^{-1/2}$ (i.e., fainter bursts will be detected as $\Delta t$
increases): the number of source counts increases as $\Delta t$, but
the square root of the background increases only as $\Delta
t^{1/2}$. However, when $T < \Delta t$---the lightcurve is a short
spike relative to the accumulation time---then the threshold flux is
proportional to $\Delta t^{1/2}$ (i.e., bursts must be brighter to
be detected for longer $\Delta t$): the number of source counts
remains constant but the square root of the background increases as
$\Delta t^{1/2}$.

Imaging is the final, determining step of the BAT's
trigger, and for short $\Delta t$ the background may not
dominate the source counts $C_s$ in the denominator of
eq.~2; the addition of $C_s$ to $B$ increases the
denominator and therefore decreases $S_i$ relative to a
simple rate trigger (eq.~1). Assume that the lightcurve is
$P A h(t;T_{90})$ where $P$ is the instantaneous peak flux
accumulated over $\Delta E$, and $A$ is the detector area.
With a maximum value of one, the lightcurve function
$h(t;T_{90})$ parameterizes the lightcurve in terms of
duration $T_{90}$. The number of burst counts accumulated
in $\Delta t_i$ is therefore $C_s = \int_0^{\Delta t_i}P A
h(t;T_{90})dt$ where I assume that $h(t;T_{90})$ peaks
within $\Delta t$ and I ignore the issue of the
registration of the burst relative to the time bin
boundaries (i.e., I assume that $\Delta t$ begins at the
beginning of the burst, and ignore the possibility that the
fluent part of the burst lightcurve straddles two time
bins). Next, let $b$ be the background rate in $\Delta E$
for the entire detector; thus the number of background
counts is $B=b\Delta t$. For my calculations I use
$b=4700$~cts~s$^{-1}$.

For a given $T_{90}$ and $\Delta t$ I calculate the threshold value
of $P_{th,r}(T_{90};\Delta t)$ for a rate trigger assuming a
threshold value of $S_r$ in eq.~1 and for an image trigger
$P_{th,i}(T_{90};\Delta t)$ assuming a threshold value of $S_i$ in
eq.~2.  Note that $S_r$ is applicable to BATSE's rate trigger, but
not to the BAT's. When there are multiple accumulation times
$\{\Delta t_i\}$, then the resulting threshold peak flux $P_{th}$ is
the minimum $P_{th}$ for the different $\Delta t_i$ values at a
given $T_{90}$. Since BATSE established a very large statistically
homogeneous burst database for $\Delta t$=1.024~s, and many burst
distributions are normalized for this value of $\Delta t$, I
normalize $P_{th}$ for different $\Delta t$ values and trigger types
to the $P_{th,r}$ for a rate trigger with $\Delta t$=1.024~s.

Figure~4a uses $h(t;T_{90})$=1 over the duration of the burst, while
Figure~4b uses $h(t)=\exp[-t/\tau]$ where $T_{90}=\tau \ln 10$. On
both figures the ratio $P_{th}(T_{90} ; \{\Delta
t_i\})/P_{th,r}(T_{90} ; \Delta t$=1.024~s) is plotted as a function
of $T_{90}$ for rate or image triggers and different sets of $\Delta
t$.  The dashed curve is for a rate trigger with BATSE's three
values of $\Delta t$=0.064, 0.256 and 1.024~s; the plotted ratio is
$P_{th,r}(T_{90} ; \{\Delta t_i\}_{\rm BATSE})/P_{th,r}(T_{90} ;
\Delta t$=1.024~s).  The decrease for durations $T_{90}$ less than
$\sim$1~s shows the increase in sensitivity to short duration bursts
that resulted from BATSE adding $\Delta t$=0.064 and 0.256~s to
$\Delta t$=1.024~s.  The solid curve shows the ratio
$P_{th,i}(T_{90} ; \{\Delta t_i\}_{\rm BAT})/P_{th,r}(T_{90} ;
\Delta t$=1.024~s) for the BAT image trigger with $\Delta t$ ranging
from 0.004~s to 26~s. As can be seen, adding $\Delta t$ values both
greater than and less than BATSE's set increases the sensitivity to
both longer and shorter duration bursts.  The increase in
sensitivity for short duration bursts is not very great because
$C_S\sim B$ (see the denominator of eq.~2); this is an unavoidable
feature of the imaging required to localize bursts.

Because burst lightcurves have very different shapes, calculating a
general detector sensitivity as a function of $T_{90}$ is very
difficult.  For example, Figures~4a and~b show that the increase in
sensitivity for long duration bursts occurs at longer durations for
the exponential lightcurves than for flat-top lightcurves.  Applying
the BAT trigger code to an ensemble of typical observed burst
lightcurves (e.g., from BATSE) would provide a better estimate of
the sensitivity as a function of duration (see Fenimore et al.
2004).

In summary, the longer $\Delta t_i$ values increase significantly
BAT's sensitivity to long duration bursts. The BAT's increase in
sensitivity to short bursts relative to BATSE is not as great
because the number of source counts becomes comparable to the number
of background counts.

\section{The Resulting Observed Burst Population}

The BAT detects mostly long duration bursts, as shown by Figure~1
(although the few short bursts have been very revealing). On average
long duration bursts are softer than short duration bursts
(Kouveliotou et al. 1993).  The BAT's detector efficiency shifts its
sensitivity to lower energies than BATSE's (Berger et al. [2005]
noted this factor), and its use of longer $\Delta t$ values
increases its sensitivity to long duration bursts. In addition, most
bursts show significant hard-to-soft spectral evolution (Ford et al.
1995), and therefore their low energy emission lasts longer; this is
an effect not considered by studying the spectral and temporal
dependencies separately. Consequently, a longer accumulation time
increases the effectiveness of a lower energy trigger band for long
duration bursts.

While the BAT's array of accumulation times increases its
sensitivity to both very short and long duration bursts relative to
BATSE's set of $\Delta t=0.064$, 0.256 and 1.024~s, the increase in
sensitivity is much greater for long bursts than for short bursts.
Whether BATSE's trigger truncated the duration distribution on the
short side was debated (e.g., Lee \& Petrosian 1996); unfortunately,
the BAT's relatively small increase in short duration sensitivity
(and its lower energy band) make it difficult to determine whether a
large population of short duration bursts exists. An analysis of
short duration rate triggers that do not result in successful image
triggers might address this issue.

Combining the results of \S 2.2 and \S 2.3, Figures~5a, b and~c show
the ratio of the BAT to BATSE flux thresholds as a function of
duration $T_{90}$ and peak energy $E_p$ for three different sets of
spectral indices.  These figures treat the energy and temporal
factors independently. Ratio values less than 1 indicate that the
BAT is more sensitive than BATSE. Also shown are a sample of BATSE
bursts for which values of $T_{90}$ and $E_p$ are available
(Mallozzi et al. 1998); the bursts in this sample provided enough
counts for spectral fits. As can be seen, the short, hard bursts are
in a region of parameter space where the BAT is less sensitive than
BATSE while the BAT is more sensitive to long, soft bursts. The
gradient of the contours shows that the BAT detects fewer short,
hard burst because its energy band is lower than BATSE's was, and
the BAT detects more long, soft bursts both because of its lower
energy band and its greater sensitivity to long bursts. This is
consistent with the shift in the duration distributions in Figure~1.
BAT's greater sensitivity to long duration bursts is consistent with
the average fluence of the {\it Swift} bursts being a factor of
$\sim2.5$ fainter than the average fluence of the BATSE bursts
(T.~Sakamoto, personal communication 2005).

When a burst occurs at high redshift, its observed spectrum
is redshifted (i.e., becomes softer) and its observed
duration is dilated.  Thus the burst is shifted towards the
parameter space region in which the BAT's sensitivity
increase is greatest.  However, evolution of the average
burst's intrinsic spectrum and duration obviously
determines where the burst began, and therefore ends, in
parameter space.

{\it BeppoSAX} and {\it HETE-II} also detect(ed) and
localize(d) bursts by forming images in low energy bands
with accumulation times longer than 1~s, and the bursts
detected by both detectors are almost exclusively long
duration bursts.  {\it BeppoSAX} formed images in the
1.8--28~keV band and {\it HETE-II} forms images in the
2--25~keV band ({\it HETE-II} also forms images in a softer
band). Thus the same factors that favor the detection of
long bursts in the BAT's burst sample are relevant to these
two detectors.

The burst detection rate depends on the spectral and temporal
sensitivities discussed in \S 2.2 and \S 2.3, respectively, and the
FOV.  The BAT's sensitivity decreases off-axis, first as a result of
area foreshortening (i.e., the detector plane is not perpendicular
to the direction to an off-axis source) and then because the outer
regions of the FOV are only partially coded. In the partially coded
region the source flux falls on only part of the detector plane, but
the entire detector plane contributes background counts.  To reduce
the dilution of source counts by background from sections of the
detector plane that are not illuminated by a source in the partially
coded region of the sky, the BAT detector plane is broken into
quadrants; each of the four quadrants, all four pairs of adjoining
quadrants and the entire detector plane are treated as independent
detectors simultaneously. Thus the burst detection sensitivity
varies across the FOV, and fainter bursts will be detected near the
center of the FOV, while only bright bursts will be detected near
the edges.  Using BATSE's burst rate as a function of peak flux
(Band 2002) and the BAT's sensitivity across its FOV, compensating
for the BAT's different energy and temporal dependencies, and
accounting for various operational factors (e.g., the deadtime
resulting from slews and SAA passages) results in an estimated burst
detection rate that is consistent with the observed detection rate
of $\sim 100$~bursts per year.

\section{Summary}

{\it Swift}'s burst sensitivity depends on the imaging performance
of the BAT, {\it Swift}'s coded mask gamma-ray detector. Using a
simplified model of the BAT's trigger, my analysis focused on the
sensitivity to bursts' temporal and spectral properties separately.
Because of the large burst database it accumulated, BATSE is the
reference detector to which I compare the BAT.

As expected from the detectors' detecting material (the BAT's CZT
vs. BATSE's NaI[{\it Tl}]), the BAT's energy band is shifted to
lower energies than BATSE's was.  Thus, for same accumulation time
(e.g., $\Delta t$=1.024~s) the BAT is less sensitive than BATSE for
$E_p>$100~keV but is more sensitive for lower $E_p$.  Note that the
relative sensitivity at fixed $\Delta t$ is only one component of
the comparison between detectors.

The BAT forms images by accumulating counts on timescales
much longer (up to 26~s) than BATSE's rate trigger (up to
1.024~s), increasing the BAT's sensitivity to long duration
bursts.  Because the number of burst counts is comparable
to the number of background counts for short bursts, the
BAT's image trigger is not as sensitive to short bursts as
a simple rate trigger would be; however, a rate trigger
would not localize the bursts.  A study of statistically
significant rate triggers of short bursts that did not
result in statistically significant point sources might
determine whether there is a large population of thus far
undetected short bursts.

The longer accumulation times increase the BAT's
sensitivity to long duration bursts, particularly for
bursts with a high level of emission over an extended
period (as opposed to long duration bursts dominated by a
short spike). The BAT detects bursts in a lower energy band
than BATSE did, and long duration bursts are softer, on
average, than short duration bursts. Consequently the BAT
detects long duration bursts preferentially.  Spectral
redshifting and time dilation of a burst's duration shift
high redshift bursts into the parameter region where BAT is
more sensitive. The same trigger characteristics explain
why {\it BeppoSAX} and {\it HETE-II} also detect(ed) long
duration bursts.

I emphasize that my semi-analytic calculations use a
simplified model of the complex BAT trigger system and my
goal is to determine how the BAT's hardware and trigger
shape the burst population the BAT detects.  My goal is
{\it not} to develop an accurate description of the
detection threshold, which is very difficult if not
impossible given the complex trigger and the time-varying
background.

\acknowledgments I thank S.~Barthelmy, E.~Fenimore, N.~Gehrels,
D.~Hullinger, H.~Krimm, D.~Palmer, A.~Parsons, T.~Sakamoto,
G.~Skinner, and J.~Tueller for their assistance and advice, and for
the preliminary data about the BAT they made available.


\clearpage

\begin{figure}
\plotone{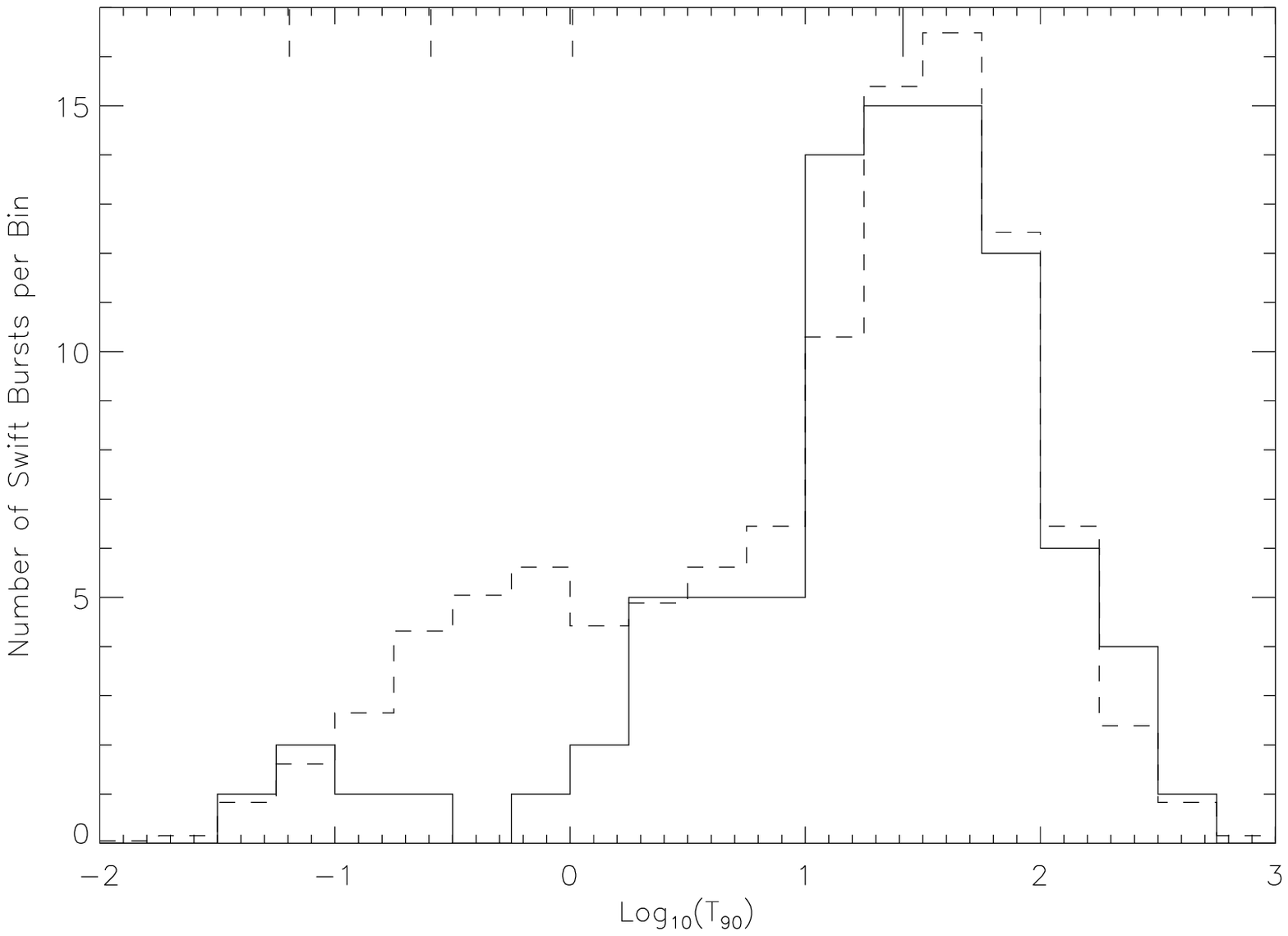}
\caption{$T_{90}$ distribution for {\it Swift} (solid) and BATSE
(dashed). An arbitrary normalization was used for the 2041 BATSE
bursts. {\it Swift} detects few of the short duration bursts that
BATSE detected.  The {\it Swift} distribution includes bursts up to
mid-December 2005.  The three short vertical dashed lines at the top
of the plot indicate the $\Delta t$ values used by the BATSE
trigger, while the solid line indicates the maximum $\Delta t$ value
used by the BAT's trigger system (the minimum value is less than the
smallest $T_{90}$ plotted).}
\end{figure}

\begin{figure}
\plotone{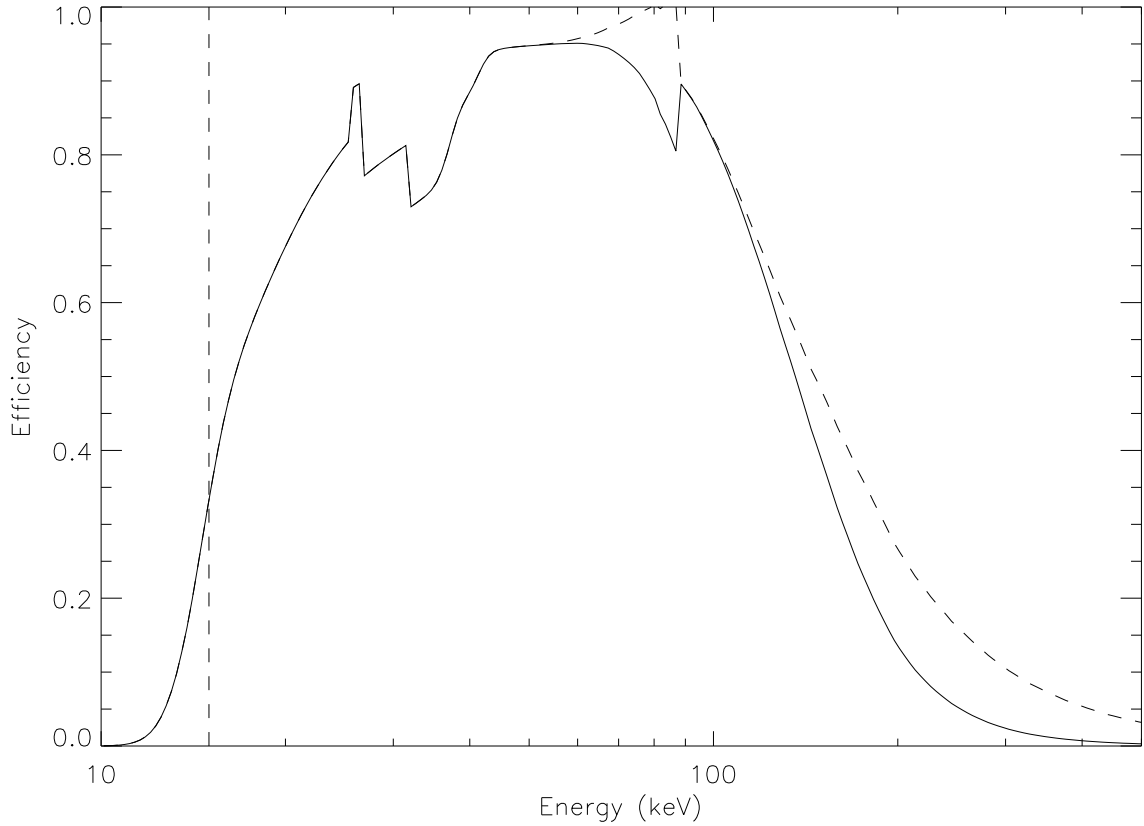}
\caption{Detector efficiency of the BAT.  Both curves are
calculated with the area of the detector plane divided by
two, accounting for the coded mask.  Relevant for imaging,
the net detector efficiency (solid curve) is the efficiency
for the {\it difference} between the fluxes through the
open and closed mask cells. The transparency of the closed
mask cells at high energy reduces the source's coded
signal.  The gross detector efficiency (dashed curve) is
the efficiency for the sum of the fluxes through the open
and closed mask cells.  In this case the mask transparency
increases the number of source photons that reach the
detector plane, increasing the gross efficiency. }
\end{figure}

\begin{figure}
\plottwo{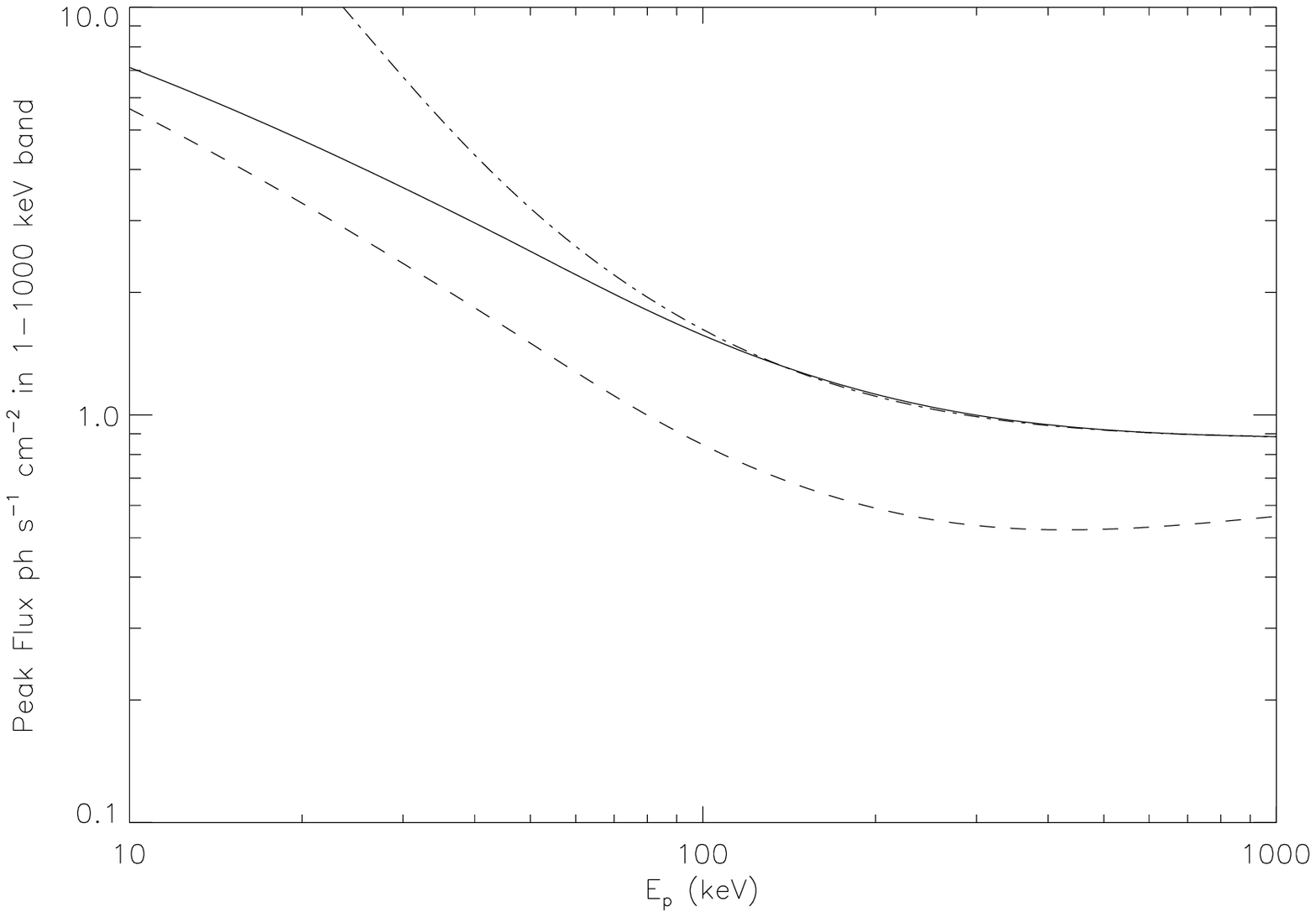}{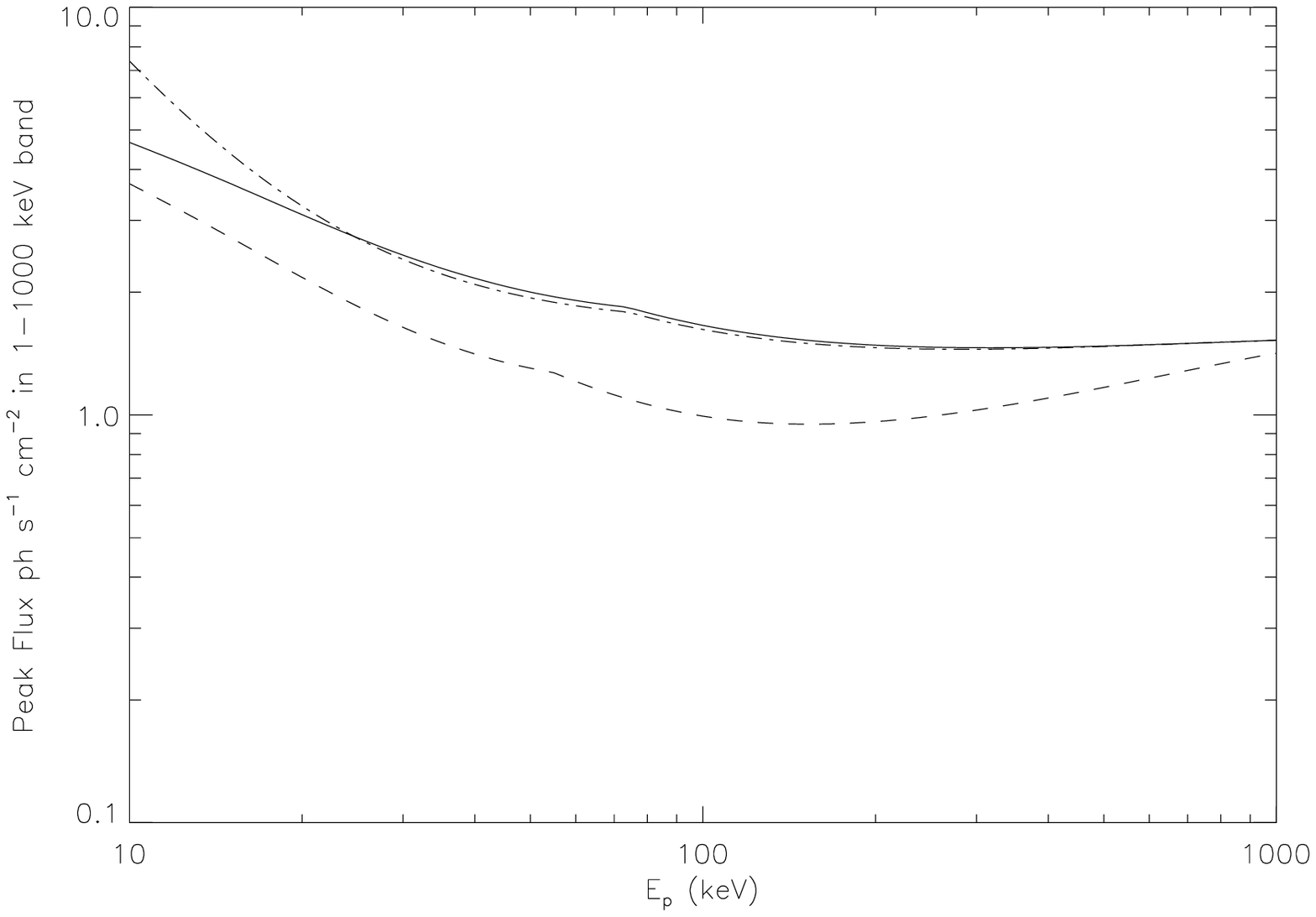}
\caption{Maximum
detection sensitivity for BATSE's LAD (left) and {\it
Swift}'s BAT (right) for $\Delta t=1.024$~s. Solid
line---$\alpha = -1$, $\beta = -2$; dashed line---$\alpha =
-0.5$, $\beta = -2$; dot-dashed line---$\alpha = -1$,
$\beta = -3$. }
\end{figure}

\clearpage

\begin{figure}
\figurenum{4} \caption{Ratio of the threshold peak flux for
a detector's set of accumulation times $\Delta t$ to the
peak flux for $\Delta t$=1.024~s as a function of the burst
duration $T_{90}$. The solid curve shows the ratio for the
BAT resulting from requiring the detection of a
statistically significant source in an image.  The dashed
curve is the ratio for BATSE's set of $\Delta t$.  Panel a:
a flat-top burst lightcurve; panel b: an exponential
lightcurve.}
\end{figure}
\plotone{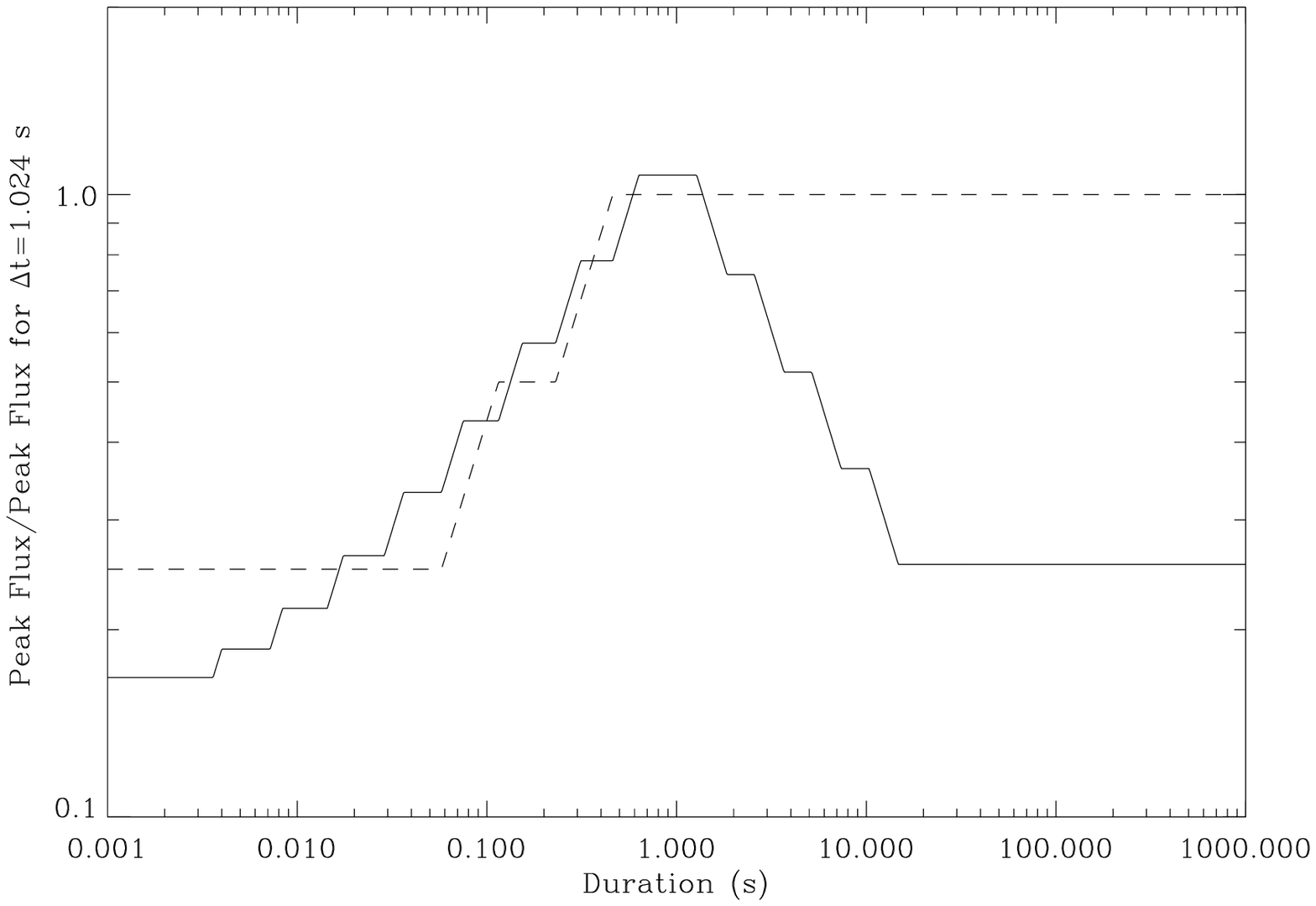}
\centerline{Fig. 4a. ---}
\clearpage
\plotone{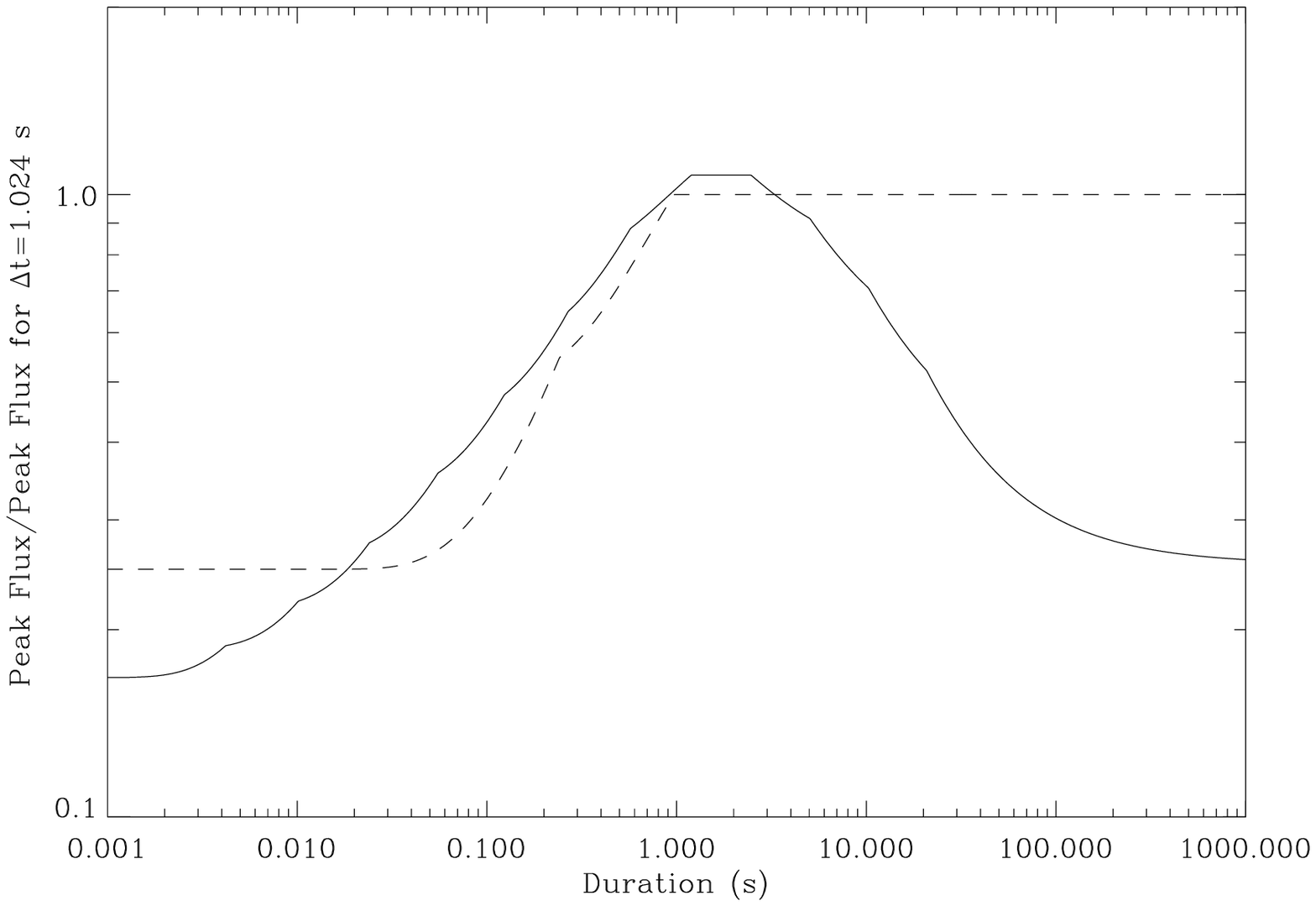}
\centerline{Fig. 4b. ---}

\clearpage

\begin{figure}
\figurenum{5} \caption{Contour plot of the ratio of the
sensitivities of the BAT and BATSE as a function of $E_p$
and $T_{90}$; a ratio less than one indicates that the BAT
is more sensitive than BATSE at that particular set of
$E_p$ and $T_{90}$. Also plotted are the $E_p$ and $T_{90}$
for a set of BATSE bursts with enough counts for spectral
fits.  The energy and temporal effects were treated
separately; differences in the burst lightcurve in
different energy bands were not considered.  Panel a:
$\alpha=$-1 and $\beta=$-2; panel b: $\alpha=$-1 and
$\beta=$-3; panel c: $\alpha=$-0.5 and $\beta=$-3. }
\end{figure}
\plotone{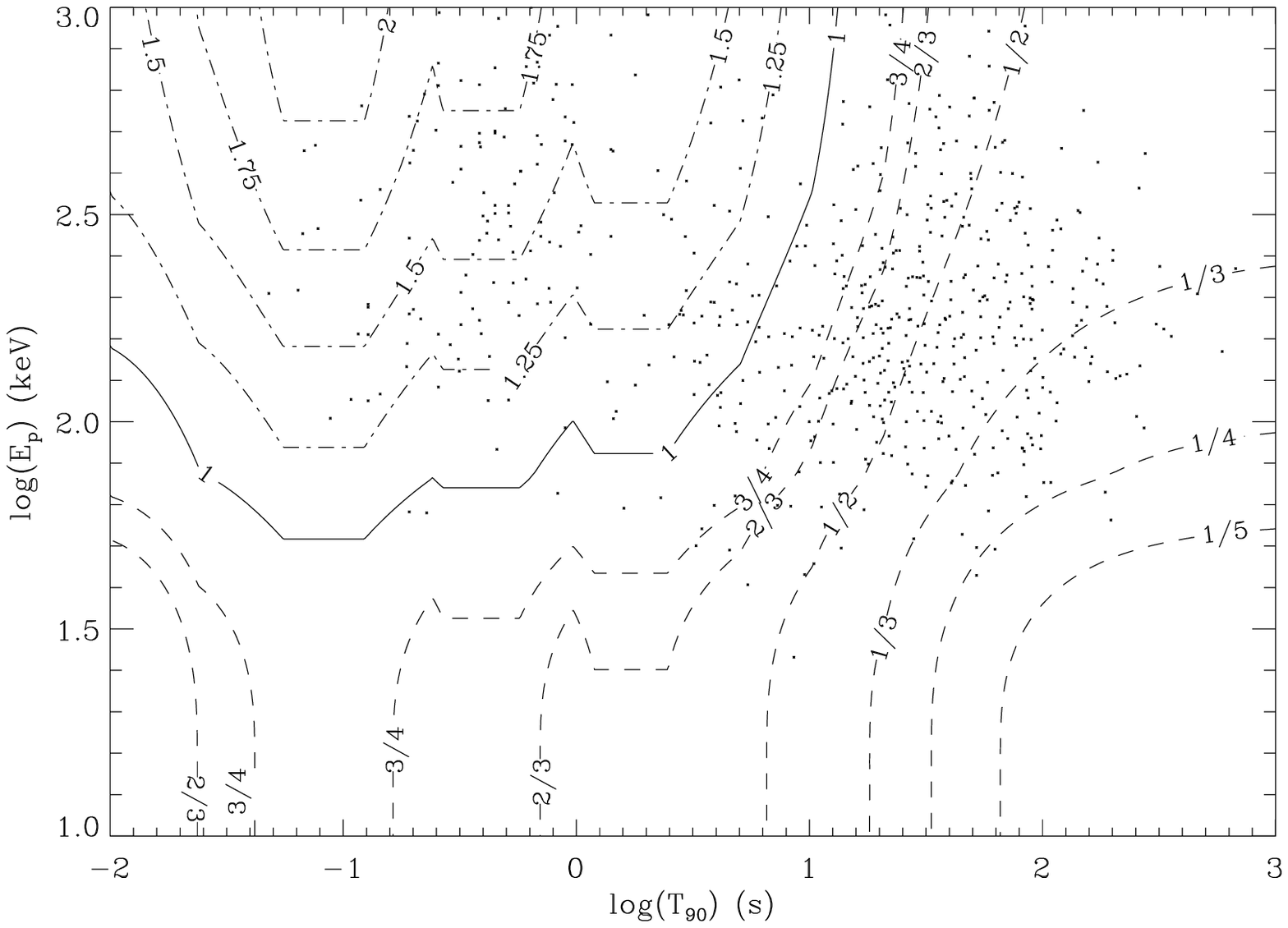}
\centerline{Fig. 5a. ---}
\clearpage
\plotone{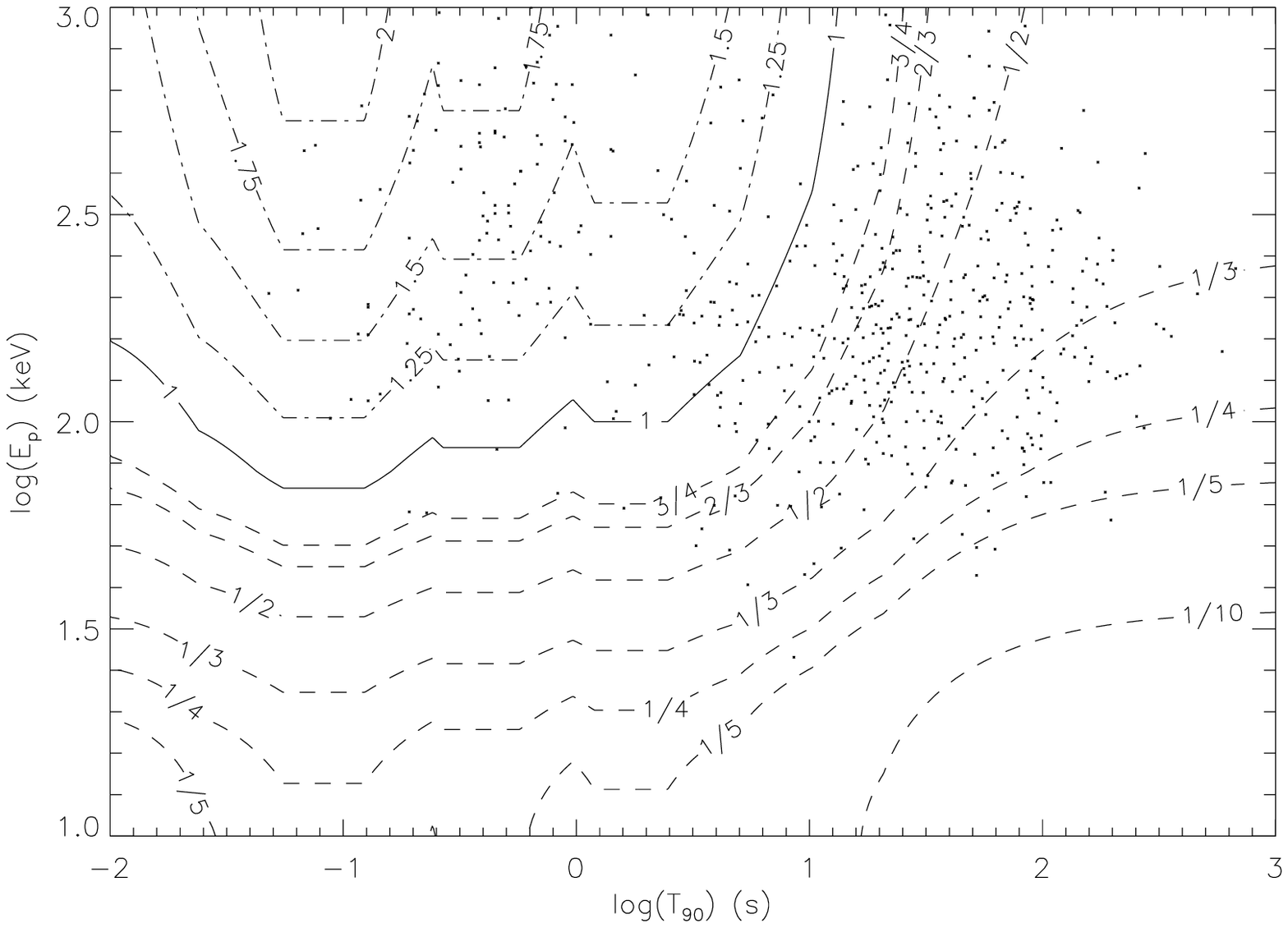}
\centerline{Fig. 5b. ---}
\clearpage
\plotone{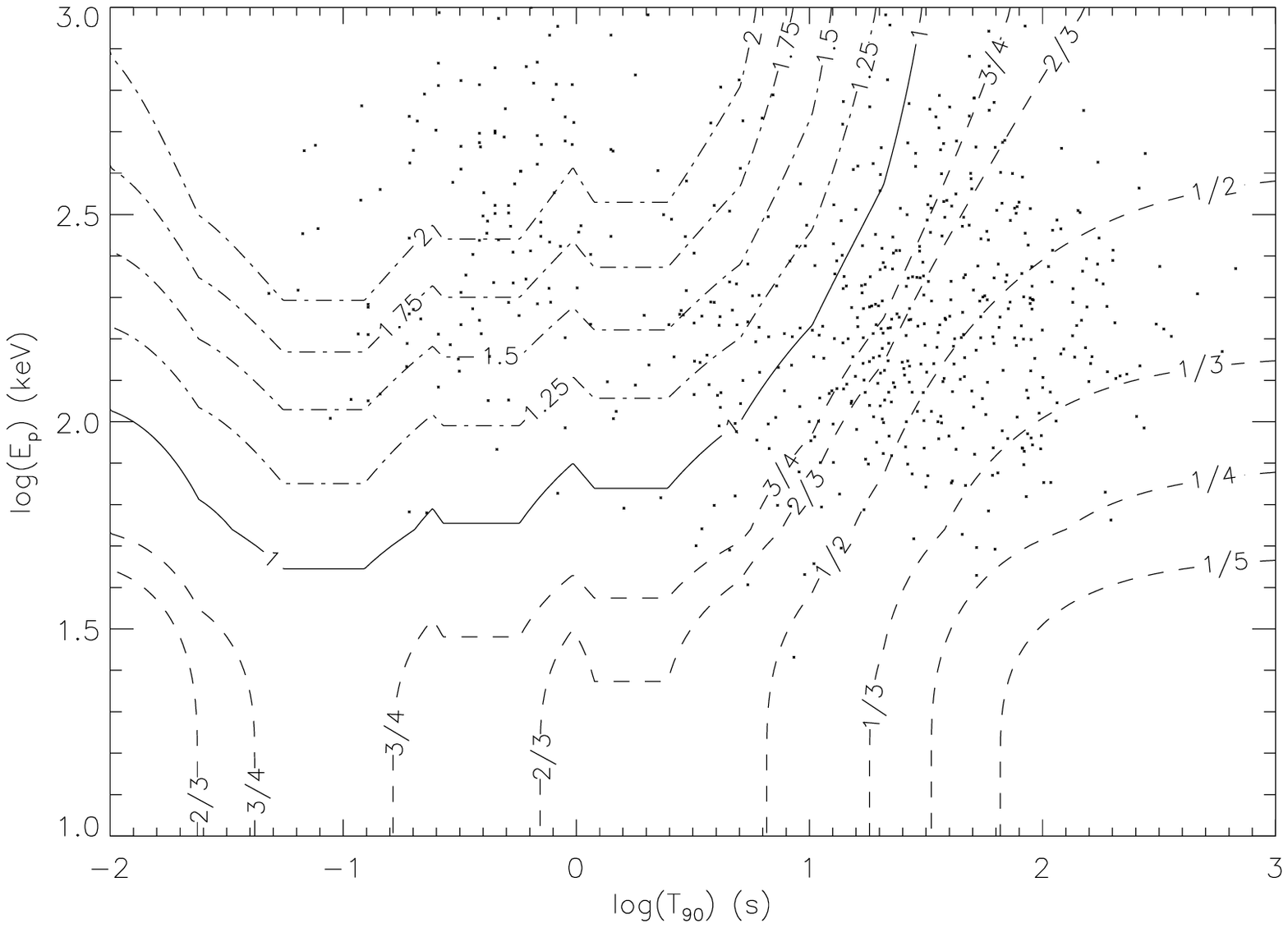}
\centerline{Fig. 5c. ---}

\end{document}